\documentclass[sigconf]{acmart}
\usepackage{url}
\usepackage{multirow}
\usepackage{balance}
\usepackage{rotating}
\usepackage{booktabs}
\usepackage{hyphsubst}
\usepackage{graphicx}
\usepackage{amsmath}
\graphicspath{{images/}}
\usepackage{times}
\usepackage{hhline}
\usepackage[caption = false]{subfig}

\usepackage{colortbl}

\usepackage{url}
\usepackage{rotating}
\usepackage{amsmath}
\usepackage{cancel}
\usepackage{amssymb}
\usepackage{enumitem}
\usepackage{pifont}

\usepackage{multirow}
\usepackage{array}
\usepackage{placeins}

\usepackage{listings}

\usepackage{boldline}

\usepackage{color}

\newcommand{\el}[1]{
		\textbf{\textcolor{olive}{EL: #1}}
}

\newcommand{\mrh}[1]{
		\textbf{\textcolor{orange}{MRH: #1}}
}

\newcolumntype{?}{!{\vrule width 1pt}}

\lstdefinestyle{json}
{
    string=[s]{"}{"},
    stringstyle=\color{black},
    comment=[l]{:},
    commentstyle=\color{blue},
}

\makeatletter

\usepackage{algorithm}
\usepackage{algorithmicx}
\usepackage{algpseudocode}
\usepackage{amsmath,amsthm,amssymb}
\usepackage{mathtools}

\DeclarePairedDelimiterX{\infdivx}[2]{(}{)}{%
  #1\;\delimsize|\delimsize|\;#2%
}

\usepackage{rotating}
\usepackage{rotating}
\usepackage{hhline}
\usepackage{hyphsubst}
\usepackage{soul}

\newlength{\Oldarrayrulewidth}

\copyrightyear{2019}
\acmYear{2019}
\setcopyright{rightsretained}
\acmConference{RecSys '19}{}{September 16, 2019, Copenhagen, Denmark}
\acmDOI{N/A}
\acmISBN{}
\acmPrice{}
\begin{document}
\title{Should we Embed? A Study on the Online Performance of Utilizing Embeddings for Real-Time Job Recommendations}

\author{Markus Reiter-Haas }
\authornote{Both authors contributed equally to this work.}
\affiliation{%
  \institution{Moshbit GmbH}
  \city{Graz, Austria} 
  }
\email{markus.reiter-haas@moshbit.com}

\author{Emanuel Lacic $^*$}
\affiliation{%
  \institution{Know-Center GmbH}
  \city{Graz, Austria} 
  }
\email{elacic@know-center.at}

\author{Tomislav Duricic}
\affiliation{%
  \institution{Graz University of Technology}
  \city{Graz, Austria} 
  }
\email{tduricic@know-center.at}

\author{Valentin Slawicek}
\affiliation{%
  \institution{Moshbit GmbH}
  \city{Graz, Austria} 
  }
\email{valentin.slawicek@moshbit.com}

\author{Elisabeth Lex}
\affiliation{%
  \institution{Graz University of Technology}
  \city{Graz, Austria} 
  }
\email{elisabeth.lex@tugraz.at}

\begin{abstract}
In this work, we present the findings of an online study, where we explore the impact of utilizing embeddings to recommend job postings under real-time constraints. On the Austrian job platform Studo Jobs, we evaluate two popular recommendation scenarios: (i) providing similar jobs and, (ii) personalizing the job postings that are shown on the homepage. Our results show that for recommending similar jobs, we achieve the best online performance in terms of Click-Through Rate when we employ embeddings based on the most recent interaction. To personalize the job postings shown on a user's homepage, however, combining embeddings based on the frequency and recency with which a user interacts with job postings results in the best online performance.



\end{abstract}

%
%


\maketitle



\begin{table*}[t!]
\centering
\renewcommand{\arraystretch}{1.25}
\scalebox{.95}{
\begin{tabular}{c?c|c|c|c|c?c|c?c|c}

& Test   & User Count & Reco Count & Days            & Approach & CTR        & $\nearrow$             & Runtime (ms) & $\searrow$           \\  \specialrule{1pt}{0pt}{0pt}

\multirow{6}{*}{\rotatebox[origin=c]{90}{\textbf{Similar Jobs}}} &
\multirow{2}{*}{Impact of embeddings} & 
\multirow{2}{*}{8,576} & 
\multirow{2}{*}{31,968} & 
\multirow{2}{*}{32} & 
 CBF   & 0.0194    & \multirow{2}{*}{18.04\%} & 51    & \multirow{2}{*}{23.53\%} \\ \cline{6-7} \cline{9-9}
&           &&&          & LAST  & \textbf{0.0229}$^{*}$ & & \textbf{39}$^{**}$   &   \\ 

                    \clineB{2-10}{2.5}

& 
\multirow{2}{*}{Influence of frequency and recency}& 
\multirow{2}{*}{4,715} & 
\multirow{2}{*}{18,464} & 
\multirow{2}{*}{15} & 
  LAST  & \textbf{0.0249}$^{**}$ & \multirow{2}{*}{75.35\%}    & \textbf{67}$^{**}$   & \multirow{2}{*}{28.72\%}  \\ 
  
\cline{6-7} \cline{9-9}
&                &&&     & BLL   & 0.0142    &   & 94   &   \\ 
                    \clineB{2-10}{2.5}

          &           
\multirow{2}{*}{Merit of recency}& 
\multirow{2}{*}{3,375} & 
\multirow{2}{*}{11,992} & 
\multirow{2}{*}{15} & 
 BLL$_{d=0.6}$  & \textbf{0.0174}$^{*}$ & \multirow{2}{*}{35.94\%}    & 97   & \multirow{2}{*}{2.06\%}  \\ \cline{6-7} \cline{9-9}
  &            &&&       & BLL$_{d=0.4}$   & 0.0128    &   & \textbf{95}   &   \\ \specialrule{1pt}{0pt}{2pt} 
  \specialrule{1pt}{0pt}{0pt}
                    
\multirow{4}{*}{\rotatebox[origin=c]{90}{\textbf{Homepage}}} &
\multirow{2}{*}{Influence of frequency and recency}& 
\multirow{2}{*}{9,620} & 
\multirow{2}{*}{26,334} & 
\multirow{2}{*}{25} & 
 BLL  & \textbf{0.0671}$^{*}$ & \multirow{2}{*}{15.69\%}    & \textbf{114}$^{**}$   & \multirow{2}{*}{13.64\%}  \\ \cline{6-7} \cline{9-9}
    &           &&&      & CF   & 0.0580    &   & 132   &   \\  
                    \clineB{2-10}{2.5}
& 
\multirow{2}{*}{Combining frequency and recency}& 
\multirow{2}{*}{9,313} & 
\multirow{2}{*}{24,907} & 
\multirow{2}{*}{19} & 
 HYB$_{BLL}$  & \textbf{0.0471}$^{**}$ & \multirow{2}{*}{33.05\%}    & 172   & \multirow{2}{*}{38.37\%}  \\ \cline{6-7} \cline{9-9}
      &         &&&      & CF   & 0.0354    &   & \textbf{106}$^{**}$   &   \\ \specialrule{1pt}{0pt}{0pt}
 
\end{tabular}
}
\renewcommand{\arraystretch}{1}
\caption{We report the mean Click-Through Rate (CTR) and the mean Runtime of the approaches utilized in the corresponding A/B tests. The increase  $\nearrow$  in accuracy and decrease  $\searrow$   in runtime is reported for the best performing approach. Moreover, we use the $*$ symbol to indicate it the results are significantly better with a p-value $< 0.05$ and the $**$ for results with a p-value $< 0.0005$.}
\label{tab:results}
\vspace{-0.7cm}
\end{table*}

\section{Introduction}
\label{sec:introduction}

Job recommender systems have become an integral part of both academia and industry for a few decades now \cite{rafter2000personalised}, which is also illustrated by the fact that XING\footnote{\url{https://www.xing.com}} has organized two recent RecSys Challenges~\cite{abel2016recsys, abel2017recsys}. In the past, research on job recommendations has mainly employed various Collaborative- and Content-Based Filtering approaches or their hybrid combinations~\cite{al2012survey,zhang2016ensemble} to improve the recommendation accuracy. Recently, learning latent item representations (i.e., embeddings) for recommender systems has become a popular technique and has shown state-of-the-art performance in the job domain. For example, the authors of \cite{yuan2016solving} use Doc2Vec~\cite{le2014distributed} to create job embeddings based on job-related content features. To test their approach, they conduct an offline evaluation, where they manually score the quality of similar jobs from a small subset of $100$ randomly selected jobs. Other works~\cite{barkan2016item2vec,lacic2017beyond,phi2016distributed} define the task at hand as an item-to-item recommendation problem and evaluate embedding approaches also in offline studies.

However, whether a user indeed accepts a recommendation can only be measured with either user studies or online evaluations. User studies are to date rarely used as they require active participation of users over a period of time~\cite{beel2013comparative} and online evaluations are expensive to set up, as they need a fully functional system with a significant userbase~\cite{campos2014time}. As a consequence, related work that reports on online studies of job recommender systems is scarce. For example, recent work~\cite{Kenthapadi2017} explored in an online study how to increase the engagement toward underserved jobs. Besides, in the case of the RecSys 2017 Challenge, the top-$25$ participating teams were allowed to publish their solutions once per day to be rolled out on the XING platform \cite{abel2017recsys}. In line with recent research~\cite{chandramouli2011streamrec,huang2015tencentrec}, recommendation approaches used in an online evaluation usually need to consider real-time constraints \cite{eksombatchai2018pixie,said2013month}, such as having response times, which are below $100$-$200$ milliseconds or immediately considering data updates in the next recommendation request.

\vspace{2mm}
\noindent 
\textbf{The present work.}
In this work, we contribute to the sparse line of research on evaluating job embeddings under real-time constraints in an online setting. For this, we learn job embeddings using the popular Doc2Vec approach. We obtain fixed-length vectors from the job description text and investigate their impact on the online performance of recommending job postings in real-time. 
Similarly, as in the offline setting of \cite{lacic2017beyond}, we further represent a user's browsing behavior by combining the extracted embeddings using a model from human memory theory, that integrates factors of frequency and recency of job posting interactions. To measure the real impact of such an approach, we perform several A/B tests on the Studo Jobs platform. That is, we compare against two popular recommendation use-cases that we tackle under the real-time constraint: (i) providing content-based recommendations for similar job postings to the one currently viewed by the user and, (ii) personalizing the homepage with job postings using collaborative filtering.
Our findings suggest that in situations when we recommend similar job postings, using embeddings based on the most recent interaction tends to improve the online performance. In contrast, combining embeddings based on the frequency and recency with which a user interacts with job postings improves the online performance when we personalize the job postings on the homepage.




\section{Recommendation Study}
\label{sec:exp}
Our study is carried out in the Studo Jobs\footnote{\url{https://studo.co/jobs}} platform. We tackle two distinguished recommendation scenarios, which the platform supports. First, we recommend similar jobs. Second, we personalize the ranked list of all possible job postings in the system to improve engagement on the homepage of Studo.

As shown in an offline study in~\cite{lacic2017beyond}, learning embeddings on the textual description of job postings can improve both the accuracy and diversity of content-based recommendations. In the present work, we learn embeddings of job postings by utilizing Doc2Vec~\cite{le2014distributed}, a variation of the widely popular Word2Vec \cite{mikolov2013distributed} approach. In order to investigate the online performance of the extracted job embeddings\footnote{We obtain the embeddings using a Doc2Vec model that we train with a window size of $20$, a learning rate of $0.025$ and $10$ negative samples.} under real-time constraints, we employ two variants for performing content-based recommendation of job postings, which are described in this section. For evaluation, we measure both Click Through Rate (CTR) and runtime. 


%
%


\vspace{2mm}
\noindent 
\textbf{Utilizing the most recent job interaction (LAST).} 
A natural way of using embeddings is to apply them in a content-based manner (e.g., \cite{musto2016learning}). That is, given a reference vector representation, the task is to find the top-$k$ similar vectors (i.e., job postings) using the Cosine similarity. As in \cite{lacic2017beyond}, to obtain this reference vector, we use the embedding of the last (i.e., most recent) job posting with which the user has interacted. With this recommendation strategy, we can study the online performance of the recommender when we recommend jobs that are similar to the one the user is currently viewing.

\begin{figure*}[ht!]
\centering
\subfloat[Impact of embeddings]{
  \centering
  \includegraphics[width=.66\columnwidth]{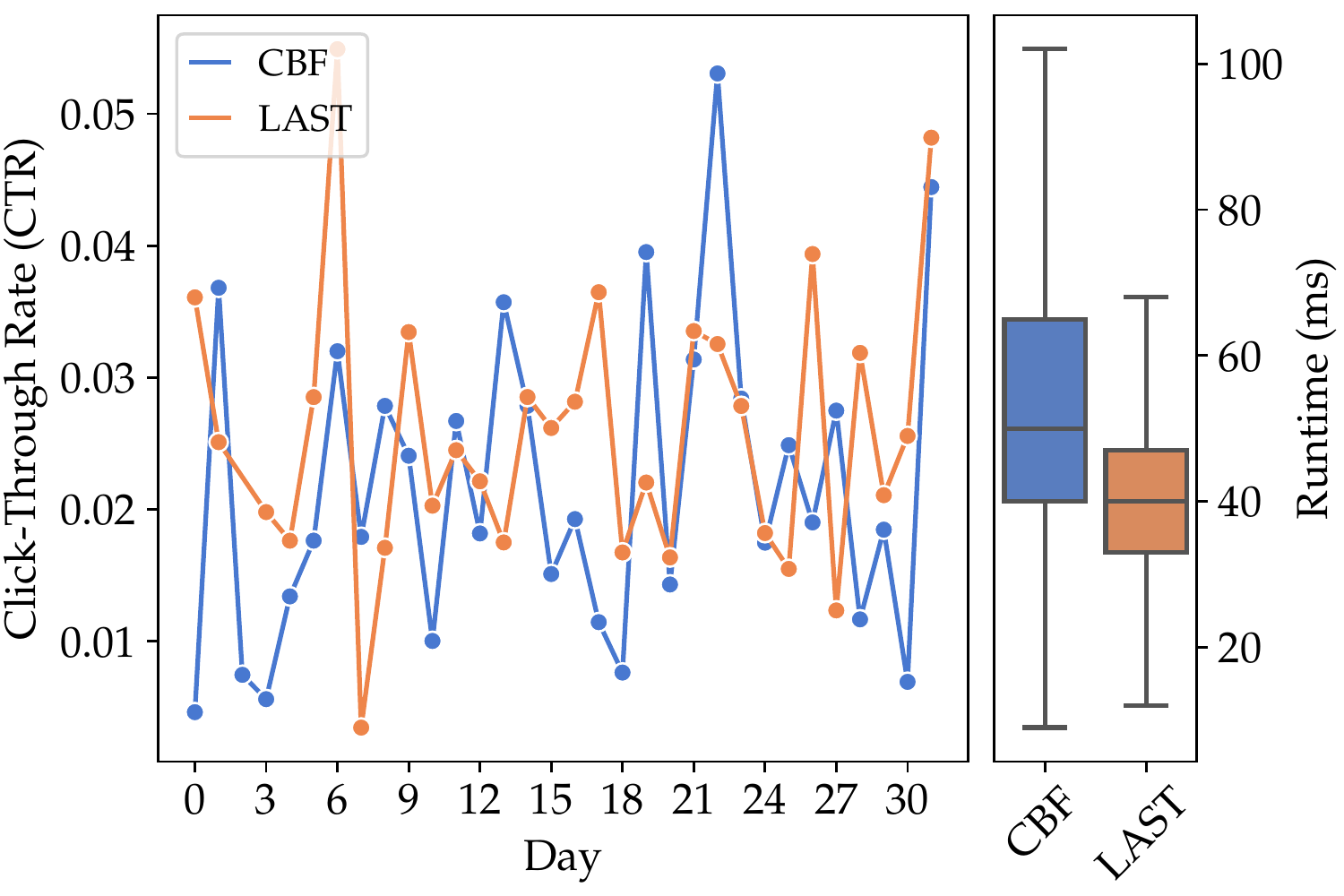}
  \label{fig:cb_vec}
}
\subfloat[Influence of frequency and recency]{
  \centering
  \includegraphics[width=.66\columnwidth]{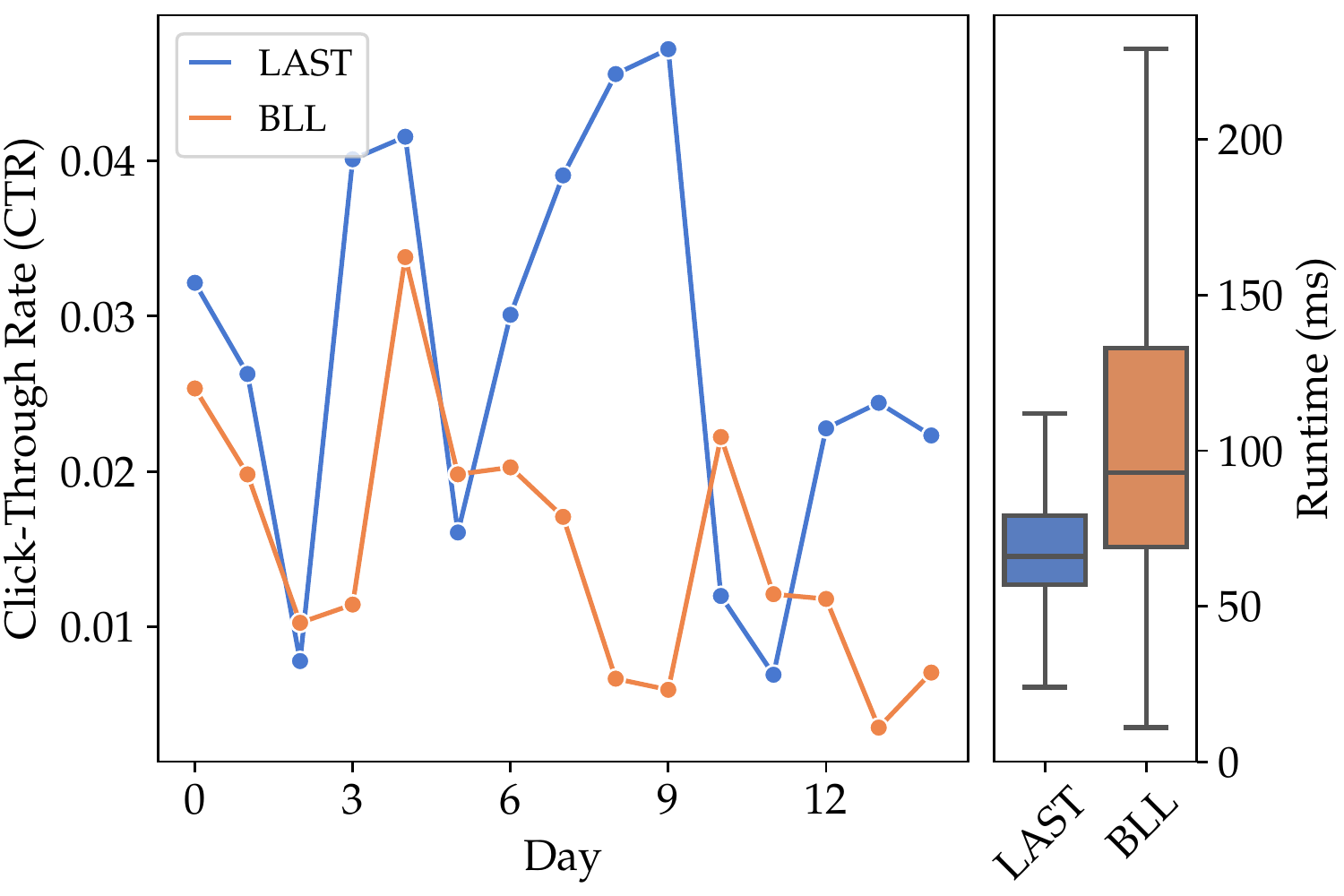}
  \label{fig:bll_vec}
} 
\subfloat[Merit of recency]{
  \centering
  \includegraphics[width=.66\columnwidth]{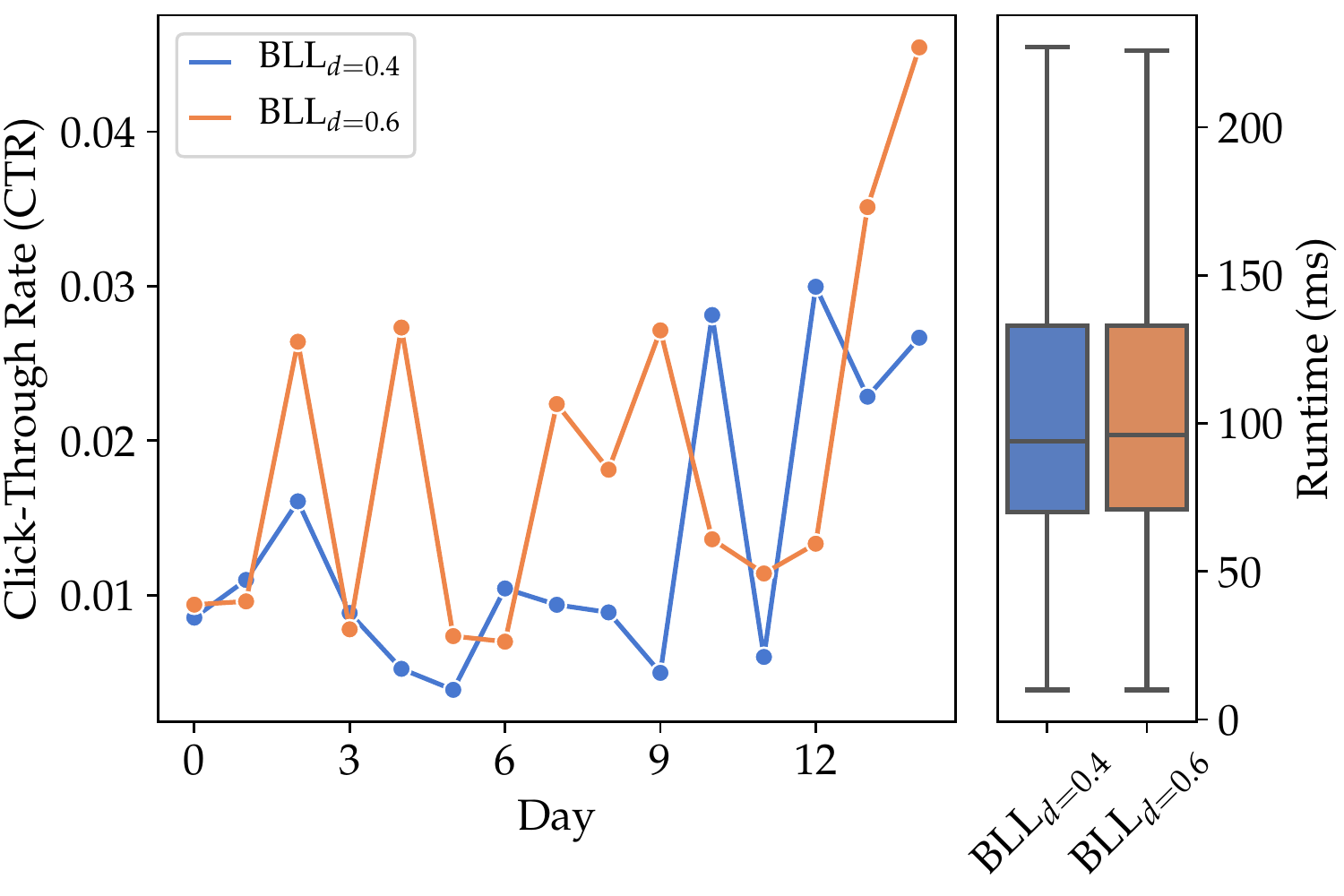}
  \label{fig:vec_bll_d_exp}
}
\caption{Analysis of incorporating job embeddings to recommend similar jobs. The reported results show a daily CTR and the distribution of the measured runtime performance.
}
\label{fig:vec_exp}
\vspace{-0.2cm}
\end{figure*}

\vspace{2mm}
\noindent 
\textbf{Integrating interaction frequency and recency (BLL).}
One issue of the previously mentioned LAST recommendation strategy is that it solely focuses on the factors of interaction recency. However, related work has shown that past interaction frequency and recency are crucial factors for personalization~\cite{lacic2017beyond,kowald2017temporal}. In this respect, the cognitive architecture ACT-R defines the Base-Level Learning (BLL) equation, which integrates these two factors to model the information access in human memory. Thus, to simultaneously account for both frequency and recency factors of job posting interactions,  we use the BLL equation to model a user's browsing behavior:
\vspace{-1.5mm}
\begin{equation}
	BLL_{u,j} = \ln(\sum\limits_{i = 1}\limits^{n}{(TS_{ref} - TS_{j,i})^{-d})}
\end{equation}
where $BLL_{u,j}$ is the BLL value for a given user $u$ and a given job $j$, and $n$ states the number of times $u$ has interacted with $j$ in the past. Moreover, $TS_{j,i}$ is the timestamp (in seconds) of when $u$ has interacted with $j$ for the $i$-th time and $TS_{ref}$ is a reference timestamp such as the time when the job recommendations are requested. The parameter $d$ is used to set the time-dependent decay of item exposure in human memory and unless stated otherwise, we set it to its default value of $0.5$ (i.e., according to Anderson et al.~\cite{anderson2004integrated}).

In this work, we use Anderson's model of human memory theory to create a reference vector representation that can be used in a content-based manner (i.e., to find similar job postings). For that, we first normalize the BLL values using a softmax function and then multiply them with the vector representations assigned to the individual job postings from a user's browsing history. This way, we form a weighted sum of embeddings based on how frequently and recently the user has interacted with the particular job postings.
As previous offline experiments have shown~\cite{lacic2017beyond}, utilizing embeddings in such a way results in a recommendation performance with lower accuracy but higher diversity when compared to the previously mentioned LAST approach.

\vspace{2mm}
\noindent 
\textbf{Adapting for real-time job recommendations.}
In practice, response times of recommendations need to be below $100$-$200$ milliseconds \cite{eksombatchai2018pixie,said2013month}. 
To adapt the LAST and BLL approaches for an online setting (i.e., to provide recommendations in real-time), we further propose to store the learned job embeddings in the form of payloads in Apache Lucene\footnote{An example of how to implement this functionality in Elastic Search, a search engine that is built on top of Apache Lucene can be found at the following link: \url{https://github.com/lior-k/fast-elasticsearch-vector-scoring}. 
}. 
Payloads are a general purpose array of bytes that are associated with a Lucene token at a particular position. Each job posting is thus annotated with multiple positions of the latent vector dimensions. The latter positional information can be used for fast retrieval and calculation of vector similarities (i.e., utilizing the Cosine similarity) at runtime. For the online study on the Studo Jobs platform, we use Apache Solr\footnote{A search engine that, similar as Elastic Search, is built on top of Apache Lucene: \url{https://lucene.apache.org/solr/}.} to store, retrieve and calculate the similarity of job embeddings in real-time.





\vspace{2mm}
\noindent 
\textbf{Experimental setup.} 
In our experiments, we measure the Click-Through Rate (CTR) and the runtime performance. To obtain the CTR, we compute the percentage of recommended job postings with which the users have interacted. For the runtime analysis, we measure the time it takes to generate each recommendation in milliseconds.
We compare two approaches at a time (i.e., conduct an A/B test) to avoid being subject to periodical changes and other anomalies. For this reason, we divide our userbase into two equal groups and assign them to one of the two approaches that we evaluate. 
We further perform a chi-squared test on the measured recommendation outcome (i.e., a user either did or did not engage with a recommendation) and a t-test on the runtime performance to determine if the differences in the reported results are statistically significant. Concerning real-time constraints, all recommendation approaches in the Studo Jobs platform calculate new recommendations for every request and filter out those job postings that the user has already interacted with in the current session.




\section{Similar Jobs}
When users view a particular job posting in the Studo Jobs platform, recommendations with similar, alternative jobs are shown to them. The location of the shown recommendations depends on the layout of the device used. On the desktop, the recommendations appear in the sidebar, while on a mobile device they will appear under the job posting description. Furthermore, this type of recommendations only suggests a short list of $3$ alternative job postings.


\vspace{2mm}
\noindent 
\textbf{Baseline: Content-Based Filtering (CBF).} A popular method in many systems for recommending similar items (i.e., jobs) is Content-Based Filtering \cite{al2012survey,pazzani2007content}. This method analyses item metadata to identify other items that could be of interest for a specific user. In Studo, this is done using TF-IDF on the description text of the job posting with which the user currently interacts. Besides being a typical pick for recommending similar items, another reason for using CBF is that it can easily be adapted for an online setting, where recommendations need to be served in real-time\footnote{As shown in \cite{lacic2014towards}, we leverage the built-in functionality of the Apache Solr search engine to recommend jobs with the most similar textual content.}.



\vspace{2mm}
\noindent 
\textbf{Impact of embeddings.}
The initial aim of this work is to investigate if utilizing embeddings, which we learn from the textual content of job postings, can outperform traditional content-based recommendations when used in a similar item scenario. For this, we first did a preliminary A/B test of the LAST approach to evaluate the impact of the embedding size. We found that in the case of Studo, having an embedding size larger than $100$ did not contribute to a higher CTR, but did increase the overall runtime performance. Table \ref{tab:results} reports on all A/B tests, for which, we use $100$ as the dimension size of job embeddings.

In Figure \ref{fig:cb_vec}, we report the performance of the LAST approach when compared to the CBF baseline. As seen, the CTR varies over the $32$-day testing period, but overall, using the job embedding from the currently viewed job posting leads to a significant increase of the CTR by $18.04\%$. Moreover, utilizing embeddings in such a way resulted in a  $23.53\%$ lower runtime  (i.e., as reported in Table \ref{tab:results}), which is a desirable effect when providing recommendations in real-time.










\vspace{2mm}
\noindent 
\textbf{Influence of frequency and recency.}
Building upon the insights on the impact of using embeddings, we evaluate the model of human memory theory during a shorter, $15$-day period. That is, we investigate if we can further enhance the recommendation performance by using the proposed BLL equation from Section \ref{sec:exp} to create the reference job vector. Interestingly enough, Figure \ref{fig:bll_vec} clearly shows that modeling a user's browsing behavior in this manner did not result in better performance than the LAST approach in terms of both CTR and runtime. As seen in Table \ref{tab:results}, we get the highest relative difference in CTR which did not justify the increased computational overhead, that resulted in higher runtime performance.

\vspace{2mm}
\noindent 
\textbf{Merit of recency.}
We hypothesize that the BLL approach did not exhibit a better performance due to the specific recommendation scenario where it was applied in (i.e., showing similar jobs to the currently viewed one). This suggests that factors of recency are especially influential in this setting and as such, we perform an additional experiment to confirm this effect.

In our previous experiment, we set the time parameter $d$ from the BLL equation to have the default value of $0.5$. However, this parameter changes the rate at which things will be "forgotten." Thus, it controls the decay of the impact of consumed items at an exponential rate. The question is therefore on whether a shorter memory (i.e., higher time decay) or a long memory is better for the setting of recommending similar job postings. For this experiment, the exponents $d=0.6$ (shorter memory) and $d=0.4$ (longer memory) were compared. As seen in Figure \ref{fig:vec_bll_d_exp} and Table \ref{tab:results}, favoring shorter memory (i.e., recency) when calculating the BLL equation resulted in a significantly better CTR. Indeed, this confirms the described effect where users expect recommendations which are similar to the more recent browsing behavior.

\section{Homepage}
As in many other systems, personalization in the Studo Jobs platform starts already on the homepage. The homepage consists of a list of $25$ job postings from which the first $5$ are the calculated recommendations. The advantage in this setting is the seamless integration of recommendations with the list of available job postings. Moreover, users not only first visit the homepage when they access the Studo Jobs portal, but also often come back to it after they stop exploring a given job posting. Such behavior results in the homepage being responsible for more than $80\%$ of all recommendations that the user has interacted with and thus suggests to be a better fit for applying the model of human memory theory to represent the user's browsing behavior.

\begin{figure}[t!]
\centering
\subfloat[Influence of frequency and recency]{
  \centering
  \includegraphics[width=.5\columnwidth]{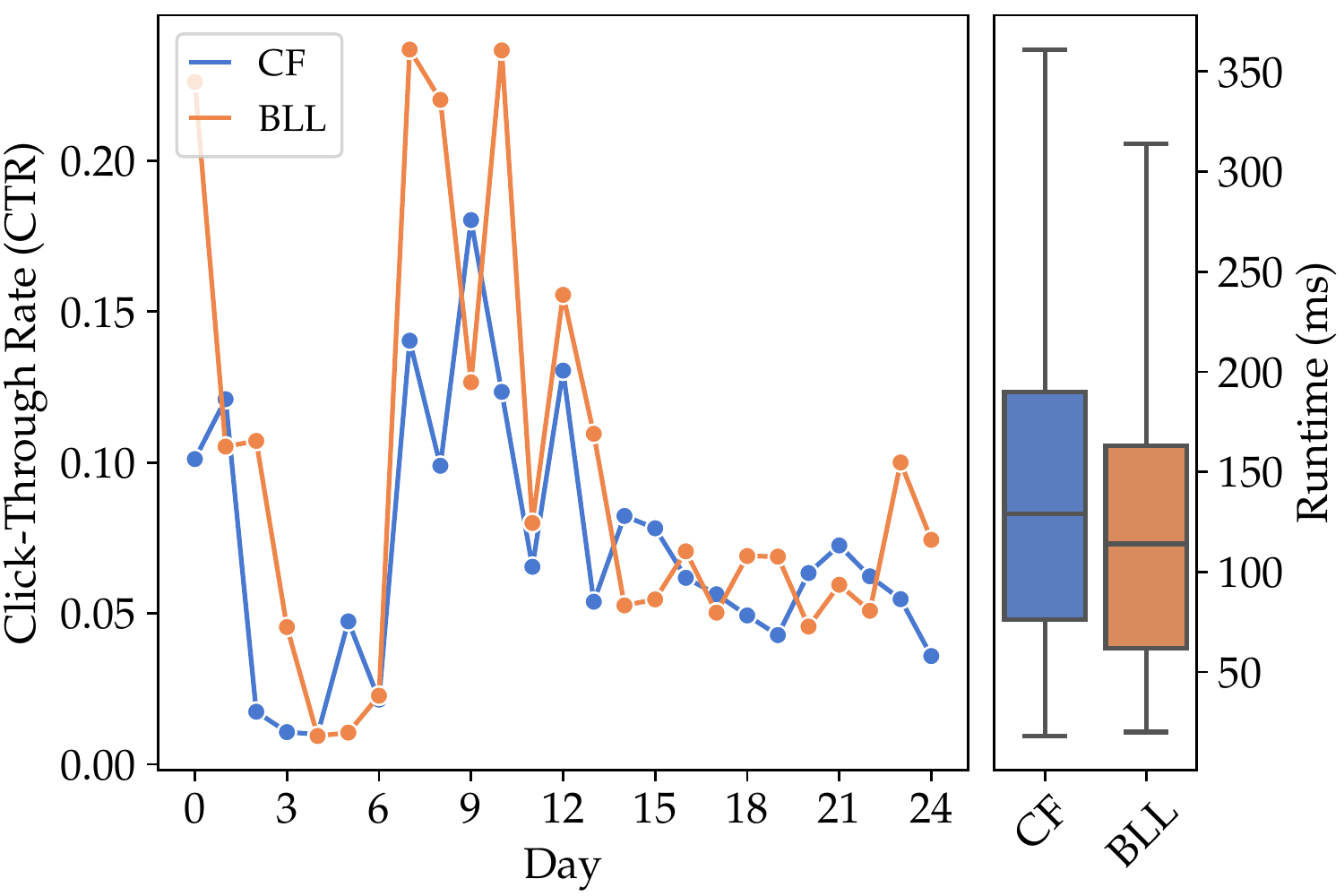}
  \label{fig:index}
}
\subfloat[Combining frequency and recency]{
  \centering
  \includegraphics[width=.5\columnwidth]{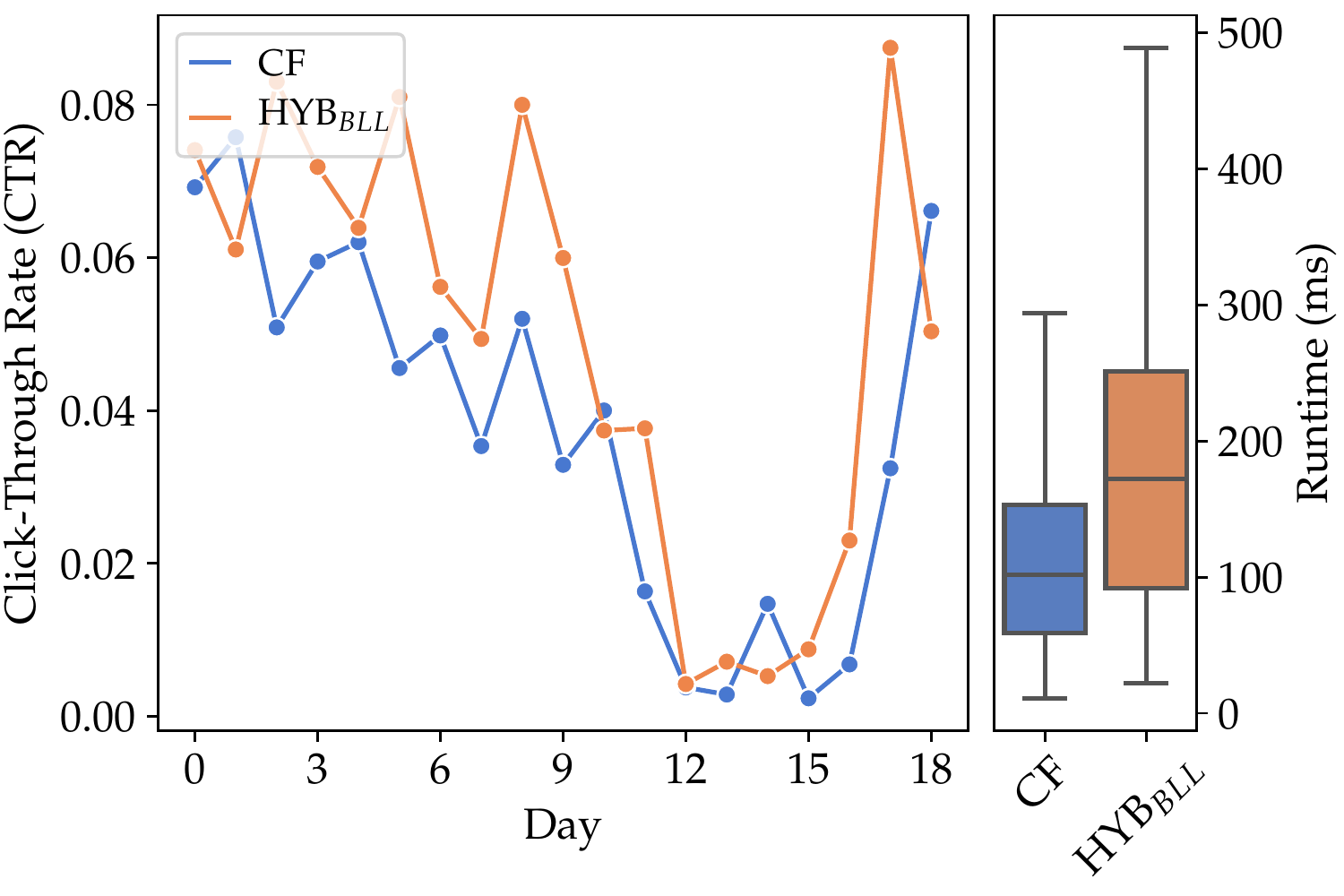}
  \label{fig:index_hybrid}
}
\caption{Online performance of job embeddings when used to personalize the homepage. 
}
\label{fig:index_exps}
\vspace{-0.2cm}
\end{figure}

\vspace{2mm}
\noindent 
\textbf{Baseline: Collaborative Filtering (CF).}
To this day, one of the most explored and utilized techniques for personalizing a system in real-time is Collaborative Filtering \cite{lacic2014towards, george2005scalable, linden2003amazon}. The Studo Jobs portal uses the User-Based Collaborative Filtering approach to personalize the job postings on the homepage. In that setting, a target user will get those job postings recommended that have been previously interacted with by similar users (i.e., the neighbors). As shown by \cite{lacic2018neighborhood}, to provide recommendations in real-time, the inverted-index structure available in the Apache Solr search engine is used to find the $k$-nearest neighbors using the Cosine similarity metric.


\vspace{2mm}
\noindent 
\textbf{Influence of frequency and recency.} 
As previously stated, we hypothesize that by incorporating the factors of frequency and recency from a user's browsing behavior, we can further enhance the online performance of recommendations on the homepage. For this, we use the BLL equation on the extracted embeddings from the user's interaction history to create a reference vector representation and recommend the top-$k$ similar job postings. To account for cold-start users, we utilize the most popular job postings as a fallback\footnote{Such a fallback strategy is used for every recommendation approach on the homepage of the Studo Jobs portal (including the CF baseline).}.

As seen in Figure \ref{fig:index} and the second part of Table \ref{tab:results}, using the BLL equation on embeddings from the user's job history manages to significantly outperform the CF baseline for both, the CTR and the runtime performance. Such results suggest that the scenario of personalizing the homepage is indeed a setting where the user expects the recommendations to consider both, factors of frequency and recency of her browsing behavior.


\vspace{2mm}
\noindent 
\textbf{Combining frequency and recency.}
Instead of replacing the CF baseline with the BLL approach, we further explore the efficacy of a hybrid combination which uses these two approaches in a round-robin fashion. We assume that for the homepage, where the user interacts most with the provided recommendations, it would make sense to allow picking from multiple sources of relevant job postings since such a recommendation strategy has often lead to the best performance in offline evaluation settings (e.g., \cite{lacic2017beyond,zhang2016ensemble}).

As seen in Figure \ref{fig:index_hybrid} and the last row of Table \ref{tab:results}, the hybrid combination of the BLL approach and the CF baseline also performs significantly better than the CF baseline concerning CTR. The relative improvement of $33.05\%$ for the CTR in this A/B test is much better than in the case when we just used the BLL approach on its own. This, however, comes with a trade-off, namely, with a significant increase in the runtime.

\section{Conclusion}
\label{sec:conc}





In this work, we contributed to the sparse line of research on evaluating job embeddings under real-time constraints in an online setting. We performed a variety of A/B tests on the Studo Jobs platform and ran evaluations concerning CTR and runtime for two different recommendation scenarios, namely, recommending similar jobs and personalizing the job postings that are shown on the homepage. 

We found that for the case of recommending similar jobs, using embeddings based on the most recent interaction provides the best online performance. 
In contrast, combining embeddings based on the frequency and recency with which a user interacts with job postings significantly improves the online performance when we personalize the jobs on the homepage.

\vspace{2mm}
\noindent 
\textbf{Limitations and Future Work.} While Doc2Vec is a popular choice for learning item embeddings, other deep learning methods such as, e.g., Autoencoders or Convolutional Neural Networks might also perform well for this task. Furthermore, we did not explore the impact of using additional user or job-related metadata on the quality of learned embeddings. We also did not study the effects of the time-dependent decay parameter $d$ from the model of human memory theory to a greater extent for personalizing the jobs shown on the homepage. As such, we aim to tackle these points as future work. Finally, the data we used for this study is proprietary and owned by Moshbit, the owner of Studo. Currently, we cannot release this data to the research community due to Moshbit's terms of service.

\balance

\bibliographystyle{abbrv}

\end{document}